\documentstyle[12pt]{article}
\begin{document}

\title{\textbf{Reconstructing interacting new agegraphic polytropic gas model in non-flat FRW universe}}
\author{K. Karami$^{1,2}$\thanks{E-mail: KKarami@uok.ac.ir} ,
A. Abdolmaleki${^1}$\thanks{E-mail:
AAbdolmaleki@uok.ac.ir}\\$^{1}$\small{Department of Physics,
University of Kurdistan, Pasdaran St., Sanandaj,
Iran}\\$^{2}$\small{Research Institute for Astronomy
$\&$ Astrophysics of Maragha (RIAAM), Maragha, Iran}\\
}

\maketitle

\begin{abstract}
We study the correspondence between the interacting new agegraphic
dark energy and the polytropic gas model of dark energy in the
non-flat FRW universe. This correspondence allows to reconstruct the
potential and the dynamics for the scalar field of the polytropic
model, which describe accelerated expansion of the universe.
\end{abstract}
\clearpage
\section{Introduction}
Type Ia supernovae observational data suggest that the universe is
dominated by two dark components containing dark matter and dark
energy (Riess et al. 1998; Perlmutter et al. 1999; de Bernardis et
al. 2000; Perlmutter et al. 2003). Dark matter (DM), a matter
without pressure, is mainly used to explain galactic curves and
large-scale structure formation, while dark energy (DE), an exotic
energy with negative pressure, is used to explain the present cosmic
accelerating expansion. However, the nature of DE is still unknown,
and people
 have proposed some candidates to describe it (for a good review see Karami et al. 2009a).

Recently, the original agegraphic dark energy (OADE) and new
agegraphic dark energy (NADE) models were proposed by Cai (2007) and
Wei and Cai (2008a), respectively. Cai (2007) proposed the OADE
model to explain the accelerated expansion of the universe, based on
the uncertainty relation of quantum mechanics as well as the
gravitational effect in general relativity. The OADE model had some
difficulties. In particular, it cannot justify the matter-dominated
era (Cai 2007). This motivated Wei and Cai (2008a) to propose the
NADE model, while the time scale is chosen to be the conformal time
instead of the age of the universe. The evolution behavior of the
NADE is very different from that of the OADE. Instead the evolution
behavior of the NADE is similar to that of the holographic DE (Cohen
et al. 1999; Horava and Minic 2000; Thomas 2002; Li 2004; Zhang and
Wu 2005; Zhang 2006; Zhang and Wu 2007; Li et al. 2009a,b;
 Sheykhi 2009b; Gao et al. 2009; Karami 2010). But some essential
differences exist between them. In particular, the NADE model is
free of the drawback concerning causality problem which exists in
the holographic DE model. The ADE models assume that the observed DE
comes from the spacetime and matter field fluctuations in the
universe (Wei and Cai 2008a, 2009). The ADE models have been studied
in ample detail by Kim et al. (2008a), Kim et al. (2008b), Wu et al.
(2008), Zhang et al. (2008), Wei and Cai (2008b), Neupane (2009),
and Sheykhi (2009a, 2010b).

Karami et al. (2009a) introduced a polytropic gas model of DE as an
alternative model to explain the accelerated expansion of the
universe. An example of a polytropic gas is a gas where the pressure
is dominated by degenerate electrons in white dwarfs or degenerate
neutrons in neutron stars. Another example is the case where
pressure and density are related adiabatically in main sequence
stars (Karami et al. 2009a).

Reconstructing the holographic and agegraphic scalar field models of
DE is one of interesting issue which has been investigated in the
literature. For instance, holographic quintom (Zhang 2006),
holographic quintessence (Zhang 2007), holographic tachyon (Zhang et
al. 2007), holographic Ricci quintom (Zhang 2009), new holographic
quintessence, tachyon, K-essence and dilaton (Granda and Oliveros
2009; Karami and Fehri 2010), interacting new agegraphic tachyon,
K-essence and dilaton (Karami et al. 2009b), and interacting
agegraphic tachyon (Sheykhi 2010a).

All mentioned in above motivate us to investigate the correspondence
between the interacting NADE and the polytropic gas model of DE in
the non-flat FRW universe. This paper is organized as follows. In
Section 2, we study the interacting NADE with the cold DM in the
non-flat FRW universe. In Section 3, we investigate the polytropic
gas model of DE. In Section 4, we suggest a correspondence between
the interacting NADE and the polytropic gas model of DE. We
reconstruct the potential and the dynamics for the scalar field of
the polytropic model, which describe accelerated expansion. Section
5 is devoted to conclusions.
\section{Interacting NADE model in non-flat FRW universe}
We consider the Friedmann-Robertson-Walker (FRW) metric for the
non-flat universe as
\begin{equation}
{\rm d}s^2=-{\rm d}t^2+a^2(t)\left(\frac{{\rm d}r^2}{1-kr^2}+r^2{\rm
d}\Omega^2\right),\label{metric}
\end{equation}
where $a$ is the cosmic scale factor and $k=0,1,-1$ represent a
flat, closed and open FRW universe, respectively. Observational
evidences support the existence of a closed universe with a small
positive curvature ($\Omega_{k}\sim 0.02$) (Bennett et al. 2003;
Spergel 2003; Tegmark et al. 2004; Seljak et al. 2006; Spergel et
al. 2007). Besides, as usually believed, an early inflation era
leads to a flat universe. This is not a necessary consequence if the
number of e-foldings is not very large (Huang and Li 2004). It is
still possible that there is a contribution to the Friedmann
equation from the spatial curvature when studying the late universe,
though much smaller than other energy components according to
observations.

For the non-flat FRW universe containing the DE and DM, the first
Friedmann equation has the following form
\begin{equation}
{\textsl{H}}^2+\frac{k}{a^2}=\frac{1}{3M_p^2}~
(\rho_{\Lambda}+\rho_{\rm m}),\label{eqfr}
\end{equation}
where $\rho_{\Lambda}$ and $\rho_{\rm m}$ are the energy density of
DE and DM, respectively. Let us define the dimensionless energy
densities as
\begin{eqnarray}
\Omega_{\rm m}&=&\frac{\rho_{\rm m}}{\rho_{\rm cr}}=\frac{\rho_{\rm
m}}{3M_p^2H^2},\\\nonumber\Omega_{\rm
\Lambda}&=&\frac{\rho_{\Lambda}}{\rho_{\rm
cr}}=\frac{\rho_{\Lambda}}{3M_p^2H^2},\\\nonumber\Omega_{k}&=&\frac{k}{a^2H^2},
\label{eqomega}
\end{eqnarray}
then, the first Friedmann equation yields
\begin{equation}
\Omega_{\rm m}+\Omega_{\Lambda}=1+\Omega_{k}.\label{eq10}
\end{equation}
Following Sheykhi (2009a), the energy density of the NADE is given
by
\begin{equation}
\rho_{\Lambda}=\frac{3{n}^2M_p^2}{\eta^2},\label{NADE}
\end{equation}
where the numerical factor 3$n^2$ is introduced to parameterize some
uncertainties, such as the species of quantum fields in the
universe, the effect of curved spacetime (since the energy density
is derived for Minkowski spacetime), and so on. The astronomical
data for the NADE gives the best-fit value (with 1$\sigma$
uncertainty) $n = 2.716_{-0.109}^{+0.111}$ (Wei and Cai 2008b). It
was found that the coincidence problem could be solved naturally in
the NADE model provided that the single model parameter $n$ is of
order unity (Wei and Cai 2008b). Also $\eta$ is conformal time of
the FRW universe, and given by
\begin{equation}
\eta=\int\frac{{\rm d}t}{a}=\int_{0}^{a}\frac{{\rm d}a}{Ha^2}.
\end{equation}
Note that in the energy density of the OADE model, the age of the
universe is appeared in Eq. (\ref{NADE}) instead of $\eta$. This
causes some difficulties. In particular it fails to describe the
matter dominated epoch properly (Cai 2007). The DE density
(\ref{NADE}) has the same form as the holographic DE, but the
conformal time stands instead of the future event horizon distance
of the universe. Thus the causality problem in the holographic DE is
avoided. Because the existence of the future event horizon requires
an eternal accelerated expansion of the universe (Wei and Cai
2008a).

From definition $\rho_{\Lambda}=3M_p^2H^2\Omega_{\Lambda}$, we get
\begin{equation}
\eta=\frac{{n}}{H\sqrt{\Omega_{\Lambda}}}.\label{eta}
\end{equation}

We consider a universe containing an interacting NADE density
$\rho_{\Lambda}$ and the cold dark matter (CDM), with $\omega_{\rm
m}=0$. The energy equations for NADE and CDM are
\begin{equation}
\dot{\rho}_{\Lambda}+3H(1+\omega_{\Lambda})\rho_{\Lambda}=-Q,\label{eqpol}
\end{equation}
\begin{equation}
\dot{\rho}_{\rm m}+3H\rho_{\rm m}=Q,\label{eqCDM}
\end{equation}
where following Kim et al. (2006), we choose
$Q=\Gamma\rho_{\Lambda}$ as an interaction term and
$\Gamma=3b^2H(\frac{1+\Omega_{k}}{\Omega_{\Lambda}})$ is the decay
rate of the NADE component into CDM with a coupling constant $b^2$.
Although this expression for the interaction term may look purely
phenomenological but different Lagrangians have been proposed in
support of it (Tsujikawa and Sami 2004). The choice of the
interaction between both components was to get a scaling solution to
the coincidence problem such that the universe approaches a
stationary stage in which the ratio of DE and DM becomes a constant
(Hu and Ling 2006). Note that choosing the $H$ in the $Q$-term is
motivated purely by mathematical simplicity. Because from the
continuity equations, the interaction term should be proportional to
a quantity with units of inverse of time. For the latter the obvious
choice is the Hubble factor $H$. The dynamics of interacting DE
models with different $Q$-classes have been studied in ample detail
by (Amendola 1999, 2000; Pavon and Zimdahl 2005; Wang et al. 2005;
Szydlowski 2006; Tsujikawa 2006; Guo et al. 2007; Caldera-Cabral et
al. 2009). It should be emphasized that this phenomenological
description has proven viable when contrasted with observations,
i.e., SNIa, CMB, large scale structure, $H(z)$, and age constraints
(Wang et al. 2006, 2007; Feng et al. 2007), and recently in galaxy
clusters (Bertolami et al. 2007, 2009; Abdalla et al. 2009).

Taking the time derivative of Eq. (\ref{NADE}), using
$\dot{\eta}=1/a$ and Eq. (\ref{eta}) yields
\begin{equation}
\dot{\rho}_{\Lambda}=-\frac{2H\sqrt{\Omega_{\Lambda}}}{{
n}a}\rho_{\Lambda}.\label{rhodot}
\end{equation}
Substituting Eq. (\ref{rhodot}) in (\ref{eqpol}), gives the equation
of state (EoS) parameter of the interacting NADE model as
\begin{eqnarray}
\omega_{\Lambda}=-1+\frac{2\sqrt{\Omega_{\Lambda}}}{3{
n}a}-b^2\Big(\frac{1+\Omega_{k}}{\Omega_{\Lambda}}\Big),\label{w}
\end{eqnarray}
which shows that in the absence of interaction between NADE and CDM,
$b^2 = 0$, $\omega_{\Lambda}$ is always larger than -1 and cannot
cross the phantom divide. However, in the presence of interaction,
$b^2\neq 0$, taking $\Omega_{\Lambda}=0.72$, $\Omega_k=0.02$
(Bennett et al. 2003), $n=2.7$ (Wei and Cai 2008b) and $a=1$ for the
present time, Eq. (\ref{w}) gives
\begin{equation}
\omega_{\Lambda}=-0.79-1.42b^2,\label{w1}
\end{equation}
which clears that the phantom EoS $\omega_{\Lambda}<-1$ can be
obtained when $b^2>0.15$ for the coupling between NADE and CDM.
\section{The polytropic gas model of DE}

Following Karami et al. (2009a), the polytropic gas equation of
state (EoS) is given by
\begin{equation}
p_{\Lambda}=K\rho_{\Lambda}^{1+\frac{1}{n}},\label{pol1}
\end{equation}
where $K$ is a positive constant and $n$ is the polytropic index. In
this model, the energy density evolves as
\begin{equation}
\rho_{\Lambda}=\left(Ba^\frac{3}{n}-K\right)^{-n},\label{pol2}
\end{equation}
where $B$ is a positive integration constant.

Using Eqs. (\ref{pol1}) and (\ref{pol2}), the EoS parameter of the
polytropic gas model of DE is obtained as
\begin{equation}
\omega_{\Lambda}=\frac{p_{\Lambda}}{\rho_{\Lambda}}=-1-\frac{Ba^{\frac{3}{n}}}{K-Ba^{\frac{3}{n}}}.\label{omegaeffpol}
\end{equation}
We see that for $K>Ba^{\frac{3}{n}}$, $\omega_{\Lambda}<-1$, which
corresponds to a universe dominated by phantom dark energy. Note
that to have $\rho_{\Lambda}>0$, from Eq. (\ref{pol2}) the
polytropic index should be even, $n=(2,4,6,\cdot\cdot\cdot)$.

Following Copeland et al. (2006), one can obtain a corresponding
potential for the polytropic gas by treating it as an ordinary
scalar field $\phi(t)$. Using Eqs. (\ref{pol1}), (\ref{pol2})
together with $\rho_{\phi}=\frac{1}{2}\dot{\phi}^2+V(\phi)$ and
$p_{\phi}=\frac{1}{2}\dot{\phi}^2-V(\phi)$, we find
\begin{equation}
\dot{\phi}^2=\frac{Ba^{\frac{3}{n}}}{\left(Ba^\frac{3}{n}-K\right)^{n+1}},\label{phidot}
\end{equation}

\begin{equation}
V(\phi)=\frac{\frac{B}{2}~a^\frac{3}{n}-K}{\Big(Ba^\frac{3}{n}-K\Big)^{n+1}}.\label{Vphi}
\end{equation}

Equation (\ref{phidot}) shows that for $K>Ba^{\frac{3}{n}}$,
$\dot{\phi}^2<0$. Therefore one can conclude that the scalar field
$\phi$ is a phantom field. Therefore, a phantom-like equation of
state can be generated from the polytropic gas DE model in a
non-flat universe.
\section{Correspondence between the interacting NADE and polytropic gas model of DE}
Here we suggest a correspondence between the interacting NADE model
with the polytropic gas model of DE in the non-flat universe. To
establish this correspondence, we compare the NADE density
(\ref{NADE}) with the corresponding polytropic gas model density
(\ref{pol2}) and also equate the EoS parameter of the interacting
NADE (\ref{w}) with the EoS parameter given by (\ref{omegaeffpol}).

Equating Eqs. (\ref{NADE}), (\ref{pol2}) and using (\ref{eta}) we
obtain
\begin{equation}
K=Ba^{\frac{3}{n}}-(3M_P^2H^2\Omega_{\Lambda})^{\frac{-1}{n}}.\label{K1}
\end{equation}
Equating Eqs.  (\ref{w}), (\ref{omegaeffpol}) and using (\ref{K1}),
we get
\begin{equation}
K=(3M_P^2H^2\Omega_{\Lambda})^{\frac{-1}{n}}\Big[-1+\frac{2\sqrt{\Omega_{\Lambda}}}{3na}-b^2\Big(\frac{1+\Omega_{k}}{\Omega_{\Lambda}}\Big)\Big].\label{K2}
\end{equation}
Substituting Eq. (\ref{K2}) in (\ref{K1}) reduces to
\begin{equation}
B=(3M_P^2H^2\Omega_{\Lambda}a^3)^{\frac{-1}{n}}\Big[\frac{2\sqrt{\Omega_{\Lambda}}}{3na}-b^2\Big(\frac{1+\Omega_{k}}{\Omega_{\Lambda}}\Big)\Big].\label{B}
\end{equation}
Now using Eqs. (\ref{K2}) and (\ref{B}), the kinetic energy term and
the scalar potential, Eqs. (\ref{phidot}) and (\ref{Vphi}), can be
rewritten as
\begin{equation}
\dot{\phi}^2=3M_P^2H^2\Omega_{\Lambda}\Big[\frac{2\sqrt{\Omega_{\Lambda}}}{3na}
-b^2\Big(\frac{1+\Omega_{k}}{\Omega_{\Lambda}}\Big)\Big],\label{phidot1}
\end{equation}
\begin{equation}
V(\phi)=3M_P^2H^2\Omega_{\Lambda}\Big[1-\frac{\sqrt{\Omega_{\Lambda}}}{3na}+\frac{b^2}{2}\Big(\frac{1+\Omega_{k}}{\Omega_{\Lambda}}\Big)\Big].\label{Vphi1}
\end{equation}
If we take $\Omega_{\Lambda}=0.72$, $\Omega_k=0.02$ (Bennett et al.
2003), $n=2.7$ (Wei and Cai 2008b) and $a=1$ for the present time,
then Eq. ({\ref{phidot1}) gives $\dot{\phi}^2<0$ when $b^2>0.15$.
This implies that the scalar field $\phi$ has a phantomic behavior.

Using definition $\dot{\phi}=\phi'H$, we can rewrite Eq.
({\ref{phidot1}) in terms of derivative with respect to $x=\ln a$ as

\begin{equation}
\phi'=
M_P\left(3{\Omega_{\Lambda}}\Big[{\frac{2\sqrt{\Omega_{\Lambda}}}{3na}
-b^2\Big(\frac{1+\Omega_{k}}{\Omega_{\Lambda}}\Big)}\Big]\right)^{1/2}.
\end{equation}
Finally, the evolutionary form of the scalar field can be obtained
as
\begin{equation}
\phi(a)-\phi(0)=M_P\int^{\ln a}_{0}
\left(3{\Omega_{\Lambda}}\Big[{\frac{2\sqrt{\Omega_{\Lambda}}}{3na}
-b^2\Big(\frac{1+\Omega_{k}}{\Omega_{\Lambda}}\Big)}\Big]\right)^{1/2}~{\rm
d}x,
\end{equation}
where we take $a_0=1$ at the present time.
\section{Conclusions}\label{II20}
Here we considered the interacting NADE model with CDM in the
non-flat FRW universe. The ADE models proposed to explain the
accelerated expansion of the universe, based on the uncertainty
relation of quantum mechanics as well as the gravitational effect in
general relativity (Cai 2007; Wei and Cai 2008a). We established a
correspondence between the NADE density and the polytropic gas model
of DE. The polytropic gas model plays a very important role in the
EoS fluid description of DE in cosmology (Karami et al. 2009a). We
reconstructed the potential of the interacting new agegraphic
polytropic as well as the dynamics of the scalar field, which
describe accelerated expansion of the universe.
\\
\\
\noindent{{\bf Acknowledgements}}\\ This work has been supported
financially by Research Institute for Astronomy $\&$ Astrophysics of
Maragha (RIAAM), Maragha, Iran.


\end{document}